\documentstyle[twocolumn,aps,prb]{revtex}
\begin{document}
\draft
\title{Pad\'e approximants for the ground-state energy of closed-shell
       quantum dots}
\author{Augusto Gonzalez\cite{adres}}
\address{Instituto de Cibernetica, Matematica y Fisica Calle E 309, Vedado,
Habana 4, Cuba\\ Universidad Nacional, Sede Medellin AA 3840, Medellin,
Colombia}
\author{Bart Partoens\cite{bart} and
Fran\c cois M. Peeters\cite{francois}}
\address{Departement Natuurkunde, Universiteit Antwerpen (UIA)
           Universiteitsplein 1, B - 2610 Antwerpen, Belgium}
\date{\today}

\maketitle

\begin{abstract}
Analytic approximations to
the ground-state energy of closed-shell quantum dots (number of
electrons from 2 to 210) are presented in the form of two-point Pad\'e
approximants. These Pad\'e approximants are constructed from the
small- and large-density limits of the energy. We estimated that 
the maximum error, reached for intermediate densities, is less than $\le 3~\%$.
Within the present approximation the ground-state is found to be unpolarized.
\end{abstract}


Quantum dots are {\it artificial atoms} which
have been a subject of intense theoretical and experimental
research in recent years.\cite{JH97} They offer the very interesting
possibility of varying the parameters characterising the dot in a broad
range, and also of fixing at will the number of electrons, $N$, confined
in the dot. External magnetic and electric fields can be added to generate new
static and dynamical effects.

For very strong confinement (high electron densities) the one-particle
Hamiltonian dominates over the coulomb repulsion, which can thus be
treated as a perturbation. On the other hand, when the confinement is
soft and the electron density becomes small, correlation effects start
to play an important role, leading at very small densities to
crystallisation,\cite{C91} as envisaged by Wigner.\cite{W34} Signals
of the crystal phase can be seen even in the level structure of the
$N=2$ - 4 systems in one dimension,\cite{JHK93} or in the ``geometry''
of the wave functions (the spatial distribution of the electron probability) of the
three-electron system in two dimensions.\cite{M96}

In the present paper, we exploit the fact that
the ground-state energy  can be computed analytically
in both the large- and small-density
limits, from which we construct an estimate for the energy for
arbitrary electron density by means
of a two-point Pad\'e approximant. The idea was applied long ago to the
three-dimensional (infinite) electron system.\cite{IM71} We have used
it recently to compute the low-lying energy levels of $N\le 5$ electrons
in a quantum dot,\cite{G97} and is extended here to dots containing
as many as 210 electrons.

One may ask why an interpolant should be a good approximation to the
ground-state energy. The reason is that the energy is a very smooth
function of the density. For small systems, it was noticed in 
Ref.~\onlinecite{JHK93}
that the low-density picture is valid up to surprisingly high densities,
and in Ref.~\onlinecite{G97} it was stressed that the regions of convergence of both
limiting expansions for the energy overlap. Below, we will argue that these 
regions also overlap for larger $N$-values.

We study quantum dots consisting of $N$ electrons moving in two
dimensions under the action of a parabolic potential. The Hamiltonian
describing the system may be written in dimensionless form as

\begin{equation}
h=\frac{H}{\hbar\omega_0}=\frac{1}{2}\sum_{i=1}^N (\vec p^{~2}+
  \vec r^{~2})+\beta^3 \sum_{i<j}\frac{1}{|\vec r_i-\vec r_j|},
\end{equation}

\noindent
where $\omega_0$ is the dot frequency, $\beta^3=\sqrt{(\frac{\mu e^4}
{\kappa^2\hbar^2})/(\hbar\omega_0)}$ is a measure for the strength of the
effective electron-electron interaction, $\mu$ is the electron effective
mass, and $\kappa$ the dielectric constant of the material the electrons are
moving in. We use the
oscillator's natural units, i.e. $\sqrt{\hbar/(\mu\omega_0)}$ for length,
$\hbar\omega_0$ for energy, etc. Notice that $\beta$ is the only
parameter entering the hamiltonian $h$. By varying $\beta$ we modify the
``density'' of the system.

Following references,\cite{G97,MP94} we will construct two-point Pad\'e
approximants from the series expansions of the energy, $\epsilon$, in
the $\beta\to 0$ and $\beta\to\infty$ limits

\begin{eqnarray}
\left. \epsilon\right|_{\beta\to 0} &=& b_0 +b_3\beta^3 +b_6\beta^6
    + \dots ,\\
\left. \epsilon\right|_{\beta\to\infty} &=& \beta^2 \{a_0 +a_2/\beta^2
    + a_4/\beta^4 + \dots\} ,
\end{eqnarray}

\noindent
in which $b_0$ is the energy of $N$ free electrons in the quadratic
potential, $b_3=<\Psi_0|\sum_{i<j} |\vec r_i-\vec r_j|^{-1}|\Psi_0>$,
$a_0$ is the classical (harmonic plus coulomb) energy of the $N$-electron
system, $a_2=\sum_i \omega_i/2$ is the zero-point energy of the classical
cluster, i.e. the $\omega_i$ are the small-oscillation frequencies around
the equilibrium configuration. $\beta\to 0$ will be called the
oscillator limit, and $\beta\to\infty$ the Wigner limit indicating that
a finite-$N$ analog of the Wigner crystal (a Wigner ``cluster'') is formed
in this limit.

We computed the coefficients $b_0$, $b_3$, $a_0$ and $a_2$
for dots with up to 210 electrons. Higher coefficients could, in
principle, be obtained although the calculations become rather involved.
From these coefficients we may construct the following approximants
interpolating between the expansions for $\beta\to 0$ and $\beta\to\infty$

\begin{eqnarray}
P_{3,2}(\beta) &=& b_0 + a_0 \beta^2 \left\{1 - \frac{1}
                {1+(b_3/a_0)\beta+(a_0/(b_0-a_2))\beta^2} \right\} ,\\
P_{4,3}(\beta) &=& b_0 + \frac{b_3 \beta^3}
                {1+q_1\beta+q_2\beta^2+q_3\beta^3}\nonumber\\
               &+& a_0\beta^2 \left\{1-\frac{1+q_1\beta}
                {1+q_1\beta+q_2\beta^2+q_3\beta^3} \right\},
\end{eqnarray}

\noindent
where $q_2=a_0/(b_0-a_2)$, $q_1=a_0 q_2/b_3$, and $q_3=(a_0 q_1-b_3)/
(b_0-a_2)$. $P_{3,2}$ and $P_{4,3}$ are the first two terms of the
second-diagonal Pad\'e sequence $\{P_{s+1,s}(\beta)\}$. The labels $s,t$
in $P_{s,t}$ mean that we are taking $s+1$ coefficients from the $\beta
\to 0$ series and $t+1$ coefficients from the $\beta\to\infty$ series
to construct the approximant. We will use (4) and (5) to estimate the
ground-state energy of the $N$-electron dot.

In what follows, we restrict the
analysis to closed-shell quantum dots, i.e. systems in which the first
$N_{shell}$ oscillator shells are filled for $\beta=0$ which will simplify our
calculation significantly. The number of
electrons is thus restricted to the following values:
$N=N_{shell}(N_{shell}+1)$. The energy in the $\beta=0$ limit is
$b_0=N(2 N_{shell}+1)/3$. For these systems, there is only one level
starting from $\epsilon=b_0$ at $\beta=0$. Its angular momentum and spin
are, respectively, $L=0$ and $S=0$. This will be the ground-state
for small $\beta$. On the other hand, for very large values of $\beta$ the
leading contributions to the energy (i.e. the coefficients $a_0$ and
$a_2$) are independent of the spin state of the system. Thus, we may
undoubtedly construct the small-$\beta$ and large-$\beta$ expansions for the
lowest rotational invariant ($L=0$), unpolarised ($S=0$) state. In the
rest of the paper, this will be called the ``ground'' state, although
it is possible that higher spin states become lower in
energy for intermediate densities.\cite{C78,K97}

Let us compute the coefficient $b_3$. The wave function for $\beta=0$
is given by the Slater determinant

\begin{equation}
\Psi_0=\left|\matrix{
              \phi_1(1)\chi_+(1)      &\dots &\phi_1(N)\chi_+(N)\cr
              \phi_1(1)\chi_-(1)      &\dots &\phi_1(N)\chi_-(N)\cr
              \dots                   &~     &~                 \cr
              \phi_{N/2}(1)\chi_+(1)  &\dots &\phi_{N/2}(N)\chi_+(N)\cr
              \phi_{N/2}(1)\chi_-(1)  &\dots &\phi_{N/2}(N)\chi_-(N)
                     }\right| ,
\end{equation}

\noindent
where we numbered sequentially the harmonic wave functions (orbitals)
$\phi_1$, \dots, $\phi_{N/2}$; and $\chi_+$, $\chi_-$ are the spin-up
and spin-down functions. The explicit form of the one-electron orbitals is

\begin{equation}
\phi_{k,l}=C_{k,|l|} r^{|l|} L_k^{|l|}(r^2)e^{-r^2/2}e^{i l \theta},
\end{equation}

\noindent
where $C_{k,|l|}=\sqrt{k!/[\pi~(k+|l|)!]}$, and 
for the composed index $n=(k,l)$ we defined the following sequential order:
\{(0,0)\}, \{(0,1), (0,-1)\}, \{(1,0), (0,2), (0,-2)\}, \dots, etc, i.e.
increasing $|l|$ inside a shell with positive-$l$ orbitals are taken first.
The energy corresponding to $\phi_{k,l}$ is $\lambda_{k,l}=1+2 k+|l|$.
In terms of these energies we have $b_0=\sum_n~\lambda_{k,l}$. The coefficient
$b_3$ can be written as

\begin{equation}
b_3=\sum_{n_1}I(n_1,n_1)+4\sum_{n_1<n_2}I(n_1,n_2)-
    2\sum_{n_1<n_2}J(n_1,n_2),
\end{equation}

\noindent
where

\begin{eqnarray}
I(n_1,n_2)&=&\int \frac{{\rm d^2}r_1 {\rm d^2}r_2}{|\vec r_1-\vec r_2|}
 |\phi_{n_1}(\vec r_1)|^2 |\phi_{n_2}(\vec r_2)|^2,\\
J(n_1,n_2)&=&\int \frac{{\rm d^2}r_1 {\rm d^2}r_2}{|\vec r_1-\vec r_2|}
 \phi_{n_1}^*(\vec r_1) \phi_{n_2}^*(\vec r_2) \phi_{n_2}(\vec r_1)
 \phi_{n_1}(\vec r_2),
\end{eqnarray}

\noindent
Both $I(n_1,n_2)$ and $J(n_1,n_2)$ can be reduced to sums of products of
Gamma functions.

The numerical results for the coefficients $b_3$ are presented in
Table~\ref{tab1}.
Notice that a simple Thomas-Fermi estimate for the present problem
\cite{LS95} shows the scaling property: $\epsilon_{TF}(N,\beta)=
N^{3/2} \epsilon_{TF}(1,N^{1/12}\beta)$ and, consequently, the $b_3$
coefficient should exhibit a $N^{7/4}$ behaviour in the large-$N$ limit
which is satisfied by the results shown in Table~\ref{tab1}.

The classical energies of the $N$-electron clusters, i.e. the
coefficients $a_0$, were computed by means of a Monte Carlo technique
with Newton optimisation.\cite{BP94,SP95} The values are also shown
in Table~\ref{tab1}. For large $N$, the Thomas-Fermi approximation predicts
the dependence $a_0\approx 1.062~N^{5/3}$. We may obtain another estimate
for $a_0$ from the classical energy of an hexagonal lattice structure,
yielding $a_0\approx 0.968~N^{5/3}$. In Table~\ref{tab1}, it is shown that the
coefficients $a_0$ of the finite systems follow the $N^{5/3}$ law when
$N\ge 30$.

Once the configuration of minimum energy of the classical problem is
found, the small-oscillation problem is solved in order to compute the
frequencies\cite{SP95} and, consequently, to find $a_2$.
The results are shown in the last column of Table~\ref{tab1}. A very
rough estimate of the zero-point energy is obtained from the dispersion
relation of the longitudinal mode of an hexagonal lattice in the (1,0)
direction, yielding $a_2\approx 0.78~N^{5/4}$. The results in 
Table~\ref{tab1} make evident the $N^{5/4}$ law for $N\ge 30$.

An evident conclusion that can be extracted from Table~\ref{tab1} is that
the properties of quantum dots with $N > 100$ are dictated by the
$N\to\infty$ asymptotics. With other words, the ground-state energy for 
such a dot should be accurately described by a $P_{4,3}$ approximant
in which $b_0\approx \frac{2}{3} N^{3/2}$, $b_3\approx 0.7~N^{7/4}$,
$a_0\approx 1.0~N^{5/3}$ and $a_2\approx 0.6~N^{5/4}$. In this sense,
these dots are more ``infinite'' than ``mesoscopic'', and their
properties will be very well reproduced by statistical theories.

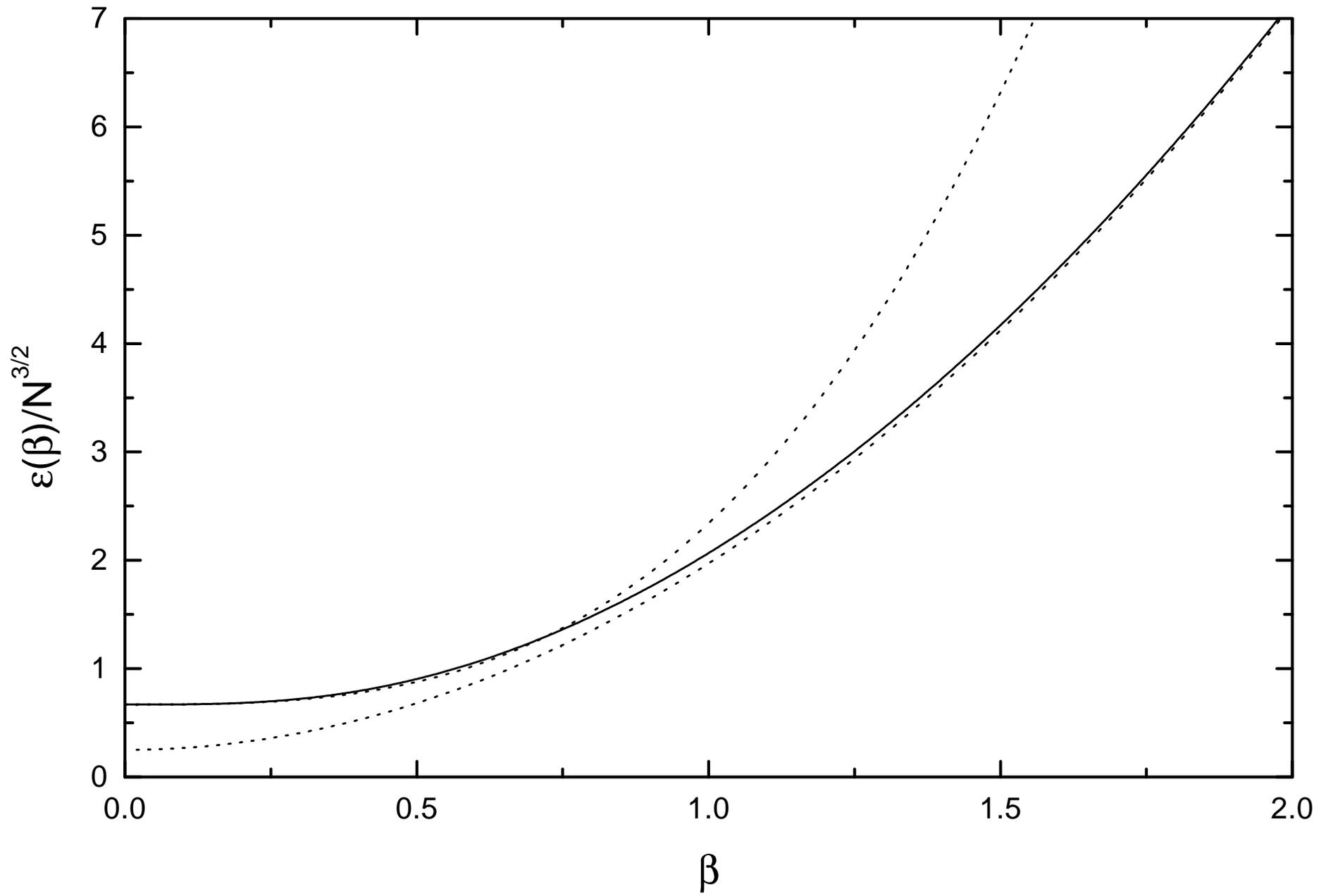
\begin{figure}
\caption{Small- and large-$\beta$ expansions (dashed curves) and
 the $P_{4,3}$ approximant (solid curve) for the ground-state of the
 42-electron system.}
\label{fig1}
\end{figure}

Another very general property following from the results shown in
Table~\ref{tab1} is the fact that the regions of convergence of the expansions
(2) and (3) overlap. As an example, we show in Fig.~\ref{fig1} the
$P_{4,3}$ approximant (solid curve) along with the small-$\beta$ and
large-$\beta$
expansions for the energy (dashed curves) for the quantum dot with
42 electrons. The asymptotic curves almost intersect one another
at $\beta\approx 0.8$. This smoothness of the dependence of $\epsilon$
on $\beta$ guarantees that an interpolant will be a very good
approximation to the actual energy at any intermediate value of $\beta$.
The same property was shown to hold\cite{G97}
for the ground-state and the excited levels of small dots with $N\le 5$.
For large dots, this seems to be a general property which
follows from the scaling laws obeyed by the coefficients. Indeed,
we can obtain a naive estimate of the radii of convergence of the series
(2) and (3) by comparing consecutive terms. The small-$\beta$ expansion
is expected to work when $\beta<\beta_0=(b_0/b_3)^{1/3}\approx
N^{-1/12}$, while the large-$\beta$ expansion will be valid for $\beta>
\beta_{\infty}=(a_2/a_0)^{1/2}\approx N^{-5/24}$. Because
$\beta_0\ge\beta_{\infty}$, the regions of convergence will overlap.

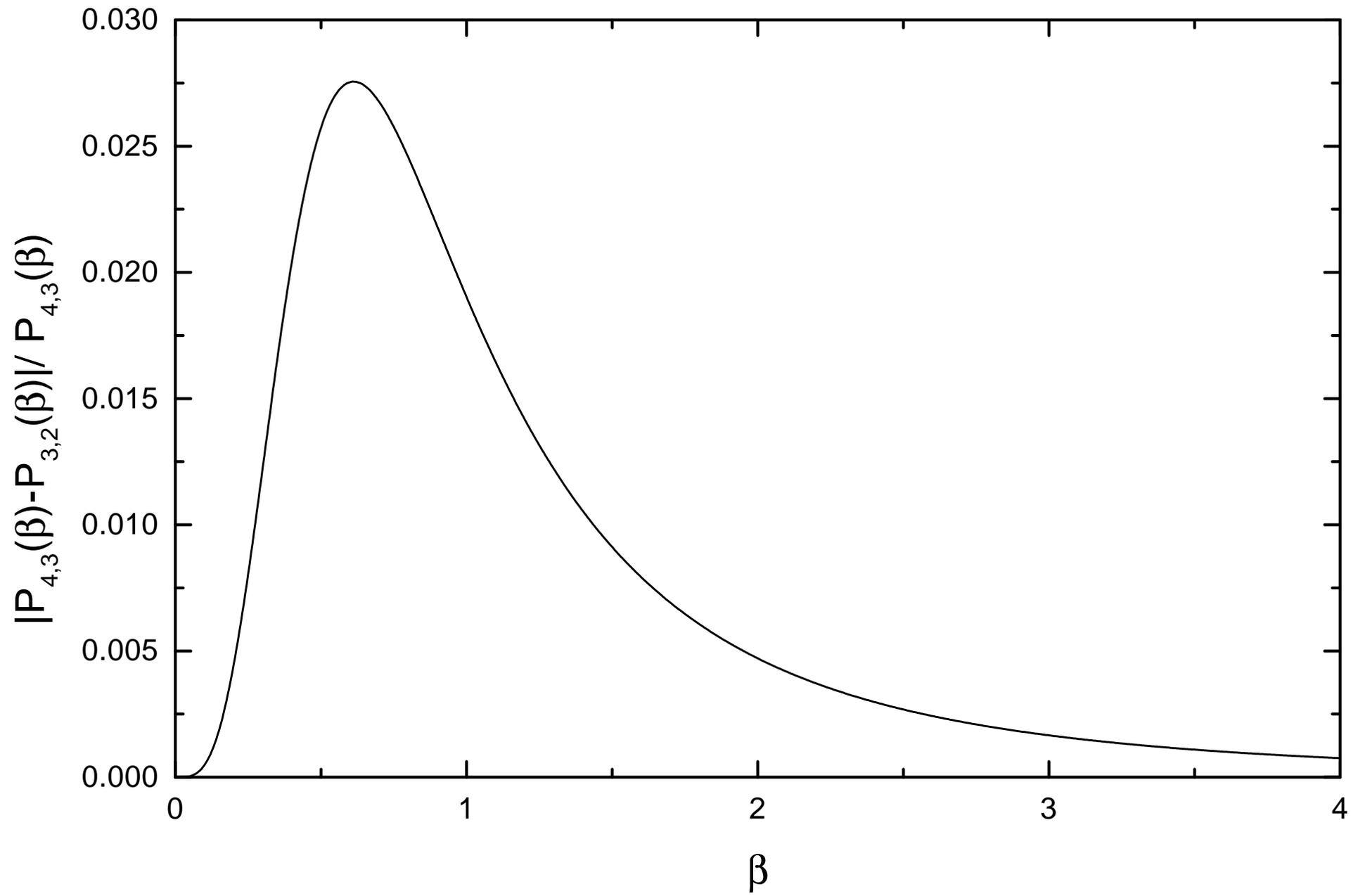
\begin{figure}
\caption{Relative difference between the $P_{3,2}$ and the
 $P_{4,3}$ approximants for the 42-electron system.}
\label{fig2}
\end{figure}

We show in Fig.~\ref{fig2} the ``convergence'' of the Pad\'e sequence, i.e.
the relative difference between the $P_{4,3}$ and the $P_{3,2}$
approximants for the 42-electron system. This figure suggests that the
relative error of the $P_{4,3}$ approximant should not exceed 3 \%
over the entire interval $0\le\beta<\infty$. The pattern is similar to
the one encountered in small dots,\cite{G97} and therefore we expect this
to be an estimate of the actual error.

Finally, we turn ourselves to the question of the absolute ground-state of
the system. In the present approximation, the energy for large $\beta$ is
independent of the polarisation state of the system. The leading approximation to the
energy is again given by Eq. (2), where $a_0$ and $a_2$ are shown in
Table~\ref{tab1}. For intermediate $\beta$ values, however, spatially antisymmetric
states may be favoured as they minimise the Coulomb repulsion. We may
easily compute the coefficients $b_0$ and $b_3$ at least for the lowest
antisymmetric state at $\beta << 1$. This state is built up by placing one
electron per orbital. If there are free orbitals in the last shell, the
occupancy of the orbitals will lead to a maximal angular momentum
(Hund's rule). For the coefficient $b_3$ we have now the
expression

\begin{equation}
b_3 = \sum_{n_1<n_2}\{I(n_1,n_2)-J(n_1,n_2)\},
\end{equation}

\noindent
where the sum runs over the occupied states.

In Table~\ref{tab2}, the angular momentum of such antisymmetric states and the
coefficients $b_0$ and $b_3$ are given as function of $N$. Pad\'e
approximants are constructed in the same way as for unpolarised states.

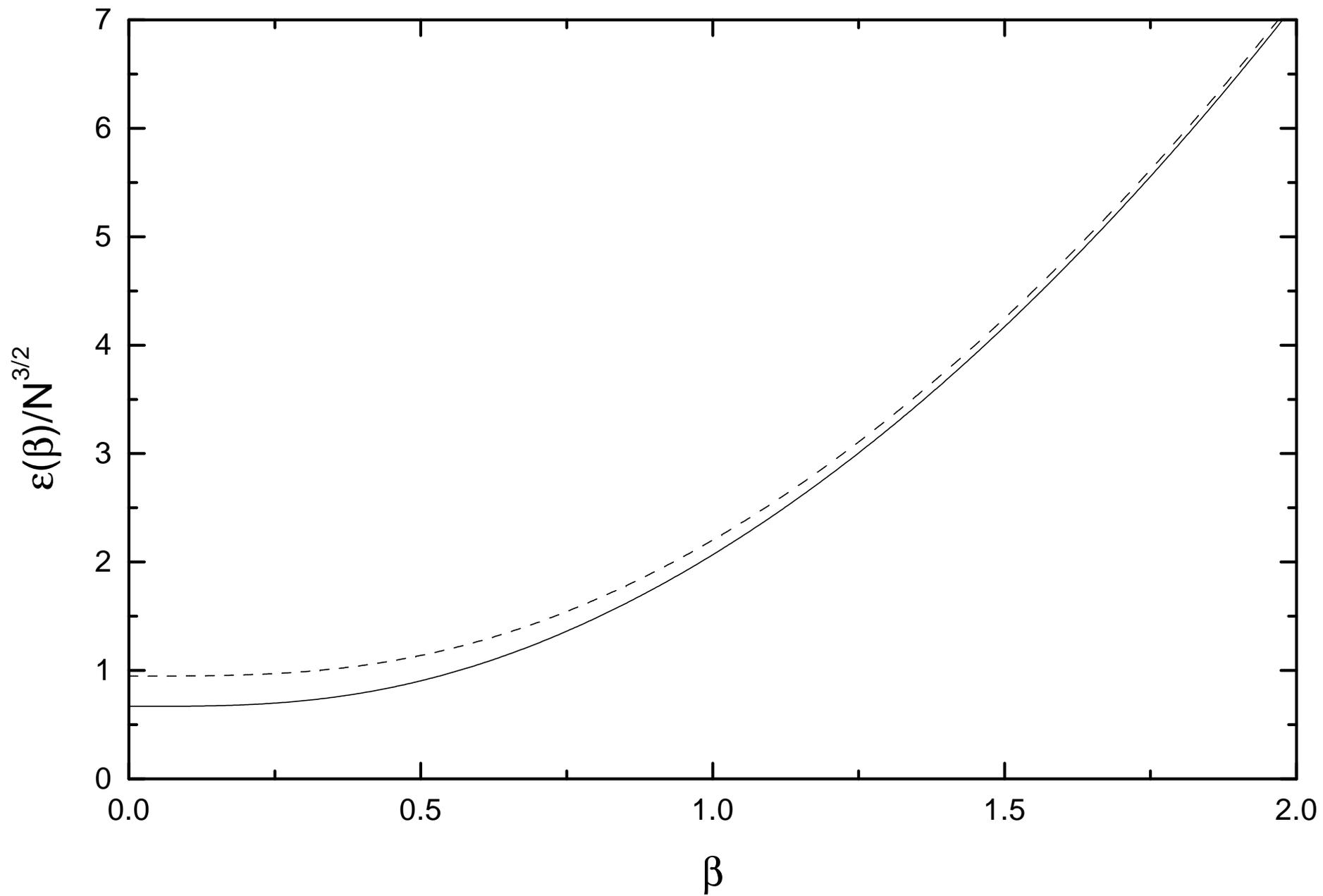
\begin{figure}
\caption{Energy estimates for the polarised (dashed curve) and 
unpolarised (solid curve) states for a quantum dot with 42 electrons.}
\label{fig3}
\end{figure}

We show in Fig.~\ref{fig3} a comparison between the estimated energies of the
polarised and unpolarised states for the dot with 42 electrons. Both
curves become very close for $\beta\ge 1$, but the unpolarised state is
always the lowest state for any $\beta$. We found that this is the case 
for all N-values shown in
Table~\ref{tab2}, which implies that in the 
present approximation the
unpolarised state corresponds to the absolute ground-state for
closed-shell quantum dots. This agrees with recent quantum mechanical density 
functional calculations.\cite{K97}

In conclusion, we have obtained Pad\'e approximants to the ground-state
energy of $N$ electrons ($2\le N\le 210$) which move in two dimensions and
are confined by a parabolic potential. The approximants are asymptotically
exact in both the high- and low-density limits. The maximum
relative error for intermediate densities is estimated to be only a few
percent of the actual energy.
This is also a typical accuracy reached in density-functional calculations
when the correlations are strong.\cite{FV94,L97}

To increase the usefulness
of the present results we were able to fit the coefficients in Table~\ref{tab1} 
as follows
\begin{eqnarray}
a_0/N^{5/3} &=& 1.062 - 0.875/N^{1/2} - 0.185/N , \\
a_2/N^{5/4} &=& 0.573 + 0.475/N^{1/2} - 0.160/N , \\
b_0/N^{3/2} &=& 2/3 + 0.083/N , \\
b_3/N^{7/4} &=& 0.731 - 0.471/N^{1/2} - 0.046/N .
\end{eqnarray}
For $N\ge 6$ the fit to all the coefficients has a relative error which is
smaller than $0.2\%$.

In principle, the Pad\'e-approximant method could equally well be applied to
other dot  configurations (i.e. non parabolic confinement,
quasi-two-dimensional systems) or to other problems such as, for example,
metallic clusters,\cite{H93,B93,CAMP95} ions in traps
,\cite{WI87,B95,K95} etc. In these cases, more numerical work
will be required to compute the coefficients $b_0$, $b_3$, $a_0$ and
$a_2$.
\vspace{1cm}

Acknowledgments: A. G. acknowledges support from the International
Centre for Theoretical Physics, Trieste, Italy, and from the Colombian
Institute for Science and Technology (COLCIENCIAS). B.P. is an aspirant,
and F.M.P. a research director with the Flemish Science Foundation
(FWO Vlaanderen). Part of this work is supported by FWO and 
the `Interuniversity Poles of
Attraction Programme - Belgian State, Prime Minister's Office - Federal Office
for Scientific, Technical and Cultural Affairs'.

\begin{table}
\caption{The coefficients $b_0$, $b_3$, $a_0$ and $a_2$ for
 closed-shell quantum dots.}
\label{tab1}
\begin{tabular}{|l|l|l|l|l|}
$N$ & $b_0/N^{3/2}$ & $b_3/N^{7/4}$ & $a_0/N^{5/3}$ & $a_2/N^{5/4}$ \\
\hline\hline
2   & 0.707107 & 0.372611 & 0.375    & 0.784567\\
6   & 0.680414 & 0.531250 & 0.674147 & 0.738995\\
12  & 0.673575 & 0.591525 & 0.793317 & 0.698756\\
20  & 0.670820 & 0.623655 & 0.856344 & 0.671474\\
30  & 0.669439 & 0.643709 & 0.895843 & 0.653428\\
42  & 0.668648 & 0.657442 & 0.922532 & 0.639116\\
56  & 0.668153 & 0.667452 & 0.941733 & 0.633269\\
72  & 0.667823 & 0.675071 & 0.956422 & 0.626248\\
90  & 0.667592 & 0.681068 & 0.967894 & 0.620456\\
110 & 0.667424 & 0.685913 & 0.977103 & 0.616908\\
132 & 0.667298 & 0.689909 & 0.984691 & 0.613269\\
156 & 0.667201 & 0.693262 & 0.991046 & 0.610032\\
182 & 0.667124 & 0.696118 & 0.996267 & 0.607227\\
210 & 0.667063 & 0.698572 & 1.00107  & 0.605239\\
\end{tabular}
\end{table}

\begin{table}
\caption{The angular momentum and the coefficients $b_0$ and $b_3$
 for the lowest spatially antisymmetric (spin polarised) state.}
\label{tab2}
\begin{tabular}{|l|l|l|l|}
$N$ & $J$ & $b_0/N^{3/2}$ & $b_3/N^{7/4}$ \\
\hline\hline
2   & 1  & 1.06066  & 0.186307\\
6   & 0  & 0.952579 & 0.378007\\
12  & 6  & 0.962250 & 0.428734\\
20  & 5  & 0.950329 & 0.486891\\
30  & 12 & 0.949386 & 0.508812\\
42  & 18 & 0.947863 & 0.525187\\
56  & 10 & 0.944959 & 0.538148\\
72  & 36 & 0.946083 & 0.544753\\
90  & 12 & 0.943998 & 0.555149\\
\end{tabular}
\end{table}

\end{document}